# Sub-nanosecond heat-based logic, writing and reset in an antiferromagnetic magnetoresistive memory

M. Surýnek[1], A. Farkaš[2,1], J. Zubáč[2,1], P. Kubaščík[1], K. Olejník[2], F. Krizek[2], L. Nádvorník[1], T. Ostatnický[1], R.P. Campion[3], V. Novák[2], T. Jungwirth[2,3], and P. Němec[1,*]

[1]Faculty of Mathematics and Physics, Charles University, Ke Karlovu 3, 121 16 Prague 2, Czech Republic

[2]Institute of Physics ASCR, v.v.i., Cukrovarnická 10, 162 53 Prague 6, Czech Republic

[3]School of Physics and Astronomy, University of Nottingham, Nottingham NG7 2RD, United Kingdom

**Thermal logic aims to create thermal counterparts to electronic circuits. In this work, we investigate experimentally the response of an analog memory device based on a thin film of an antiferromagnetic metal CuMnAs to bursts of heat pulses generated by the absorption of femtosecond laser pulses at room ambient temperature. When a threshold temperature in the heat-based short-term memory of the device is exceeded, the output of the in-memory logic operations is transferred within the same device to a long-term memory, where it can be retrieved at macroscopic times. The long-term memory is based on magnetoresistive switching from a reference low-resistive uniform magnetic state to high-resistive metastable nanofragmented magnetic states. The in-memory heat-based logic operations and the conversion of the outputs into the electrically-readable long-term magnetoresistive memory were performed at sub-nanosecond time scales, making them compatible with the GHz frequencies of standard electronics. Finally, we demonstrate the possibility of rapidly resetting the long-term memory to the reference low-resistive state by heat pulses.**

**Keywords**

Analog antiferromagnetic memory device, Ultrafast optical switching, Short-term to long-term memory transfer, Thermal logic

[*] Electronic mail: petr.nemec@matfyz.cuni.cz

## Introduction

In the conventional von Neumann computer architecture, digital data processing and memory are carried out in spatially separated functional sections. Nowadays, the ever-increasing demand for computing power is complemented by the demand for higher efficiency in information technology (IT), especially due to the boom of artificial intelligence-based applications. This requires a change from traditional to alternative computational schemes, with spintronics [1-3] and bio-inspired neuromorphic computing [4-6] as examples of the most commonly pursued alternative approaches. Moreover, even the most efficient digital memory bits in current computing architectures are intrinsically dissipative, generating heat that cannot be decreased below the Landauer fundamental limit [7]. Heat and its dissipation are generally considered a nuisance in IT [8-10] as they limit the processor frequency [11], for example. The only exception is heat-assisted magnetic recording (HAMR) technology [12], which is used in the most advanced hard-disk drive (HDD) technology. Here, the temperature-induced decrease of magnetic anisotropy in granular recording media enables to write bits with moderate magnetic fields while maintaining the required long-term stability of the stored data [12], leading to a production of high storage-density HDDs. While in HAMR heat plays only a passive role, there are several proposals that heat can also play an active role in IT. One possible approach is thermal logic [13-15], which aims to create thermal counterparts to electronic circuits using thermal logic gates [13,16,17], diodes [18-20], transistors [21-23], and memories [24-26]. Since the calculations are performed by driving heat currents through a heat circuit using a temperature gradient, they can be also driven by the excess heat in standard integrated circuits. The aim is to perform further logical operations by recycling some of the heat already generated in a computation architecture [13,14]. Consequently, hybrid electrical/thermal systems, where some parts of the circuit perform electrical and other parts perform thermal computations, are a potentially fruitful platform for energy-efficient computing. To enable their mutual compatibility, the thermal logical values in heat circuits must eventually be converted to electrical ones. This can be achieved by complementing the heat-logic and heat-memory units with a heat-to-charge conversion unit. For this purpose, temperature-biased normal-metal/ferromagnetic-insulator/superconductor tunnel junctions can be used, for example [13]. To be compatible with the GHz frequencies of standard electronics, the heat-based logic operations should be performed at sub-nanosecond timescales. However, the characteristic switching times in the experimentally achieved realizations of components for the heat-circuits are generally much longer, e.g. 500 ms to minutes for thermal transistors [21-23].

Antiferromagnets (AF) are promising materials for spintronic applications as they combine several favorable properties [27-29]. For example, the absence of net magnetization and stray fields eliminates crosstalk between adjacent devices, enabling their high-density placement and making them robust against external magnetic fields. Moreover, due to the strong exchange coupling between the sublattices in AF, the intrinsic resonance frequencies are in the terahertz range, as opposed to gigahertz frequencies in ferromagnets, which promises the corresponding orders of magnitude faster device operation [30-32]. At the same time, the presence of metastable non-uniform magnetic states of a variable resistivity [33-35] also makes AF potentially suitable for bio-inspired neuromorphic computing [36-38].

Metallic collinear antiferromagnetic CuMnAs is a material where several pioneering experiments related to switching of AF have been performed [31,35,39], including the first experimental realization of all-electrical room-temperature antiferromagnetic memory devices operated by a PC via a common USB interface [34]. In this material, two distinct mechanisms can be used for switching of the magnetic order. At current densities of $\sim 10^6$ $Acm^{-2}$, a reorientation of the Néel vector by a field-like spin-orbit torque is observed that can be detected by anisotropic magnetoresistance (AMR), with relative resistance changes limited to tenths of a percent [34,39]. At higher current densities of $\sim 10^7$ $Acm^{-2}$, switching of CuMnAs into highly resistive states occurs, with the relative resistivity change reaching 20% at room temperature and approaching 100% at low temperatures [35]. This effect is termed quench switching and occurs when the system is heated to the vicinity of the Néel temperature by the switching pulse, followed by a rapid quenching to a metastable magnetically disordered state [35]. Previous studies attributed these higher-resistance states to the formation of unconventional nanoscale magnetic textures in the AF [35]. In contrast to the spin-orbit torque reorientation of the Néel vector, which is controlled by the angle or polarity of the switching current [34,39,40], the quench-switching mechanism is independent of the direction of the writing current and can be triggered by electrical, THz-field or optical-laser pulses [31,35]. The only requirement is that the material temperature exceeds a threshold temperature that scales with the Néel temperature [41], which is $\approx 453$ K for CuMnAs.

In our earlier work [42], we extended the research on CuMnAs-based memory devices to the area of in-memory analog computing, which is a non-von Neumann computing approach where certain computational tasks are performed in the memory itself [43-45]. In particular, we experimentally separated the heat-related and quench-switching-related resistance signal dynamics induced by a single femtosecond laser pulse in a CuMnAs-based memory device [42].

We proposed that heat-related dynamics on a time scale of picoseconds to hundreds of nanoseconds can be used as a short-term memory (STM) in which information about input stimuli, represented by laser pulses, is temporarily stored. When the threshold for quench-switching is reached, the information is transferred to the device's variable resistance, serving as a magnetic-based long-term memory (LTM) with relaxation comprising two prominent components with time constants ~ 10 ms and ~ 10 s. Finally, we discussed the potentially attractive properties of these devices for several research areas, including bio-inspired neuromorphic devices and heat-based logic [42,46].

In this work, we investigate the possibility of combining approaches from these two areas. In particular, we study CuMnAs-based memory devices that integrate the functionalities of a heat-based logic-in-memory and a heat-to-charge conversion unit. We experimentally demonstrate the applicability of the psychological model of the human brain's process of remembering and forgetting, which is often applied in artificial synaptic devices [47-49], to heat-based information encoding. We show that the time scale of heat-based operations, namely the logic operations and the transfer to the electrically-readable long-term memory within the same device, can be as short as (sub-)nanoseconds. We also demonstrate the possibility of rapidly resetting the device's long-term memory from the metastable high-resistive state to the reference low-resistive state by applying heat pulses.

## Results

### Similarity between CuMnAs-based analog memory device and biological synapse

In the human brain, data processing is performed by the computing-in-memory architecture. At the cellular level, memories are stored by modifying the synaptic strengths, which can be increased or decreased, depending on previous activity patterns – the very active synapses are likely to become stronger, and the less active ones tend to become weaker [50,51]. Synaptic activity can be divided into short-term plasticity [51-54] and long-term plasticity [55-57]. Long-term potentiation (i.e., an increase in synaptic strength) and long-term depression (a decrease in synaptic strength), which occur on a time scale of seconds or more, are believed to form the basis of learning and memory. In contrast, short-term plasticity, which occurs on a millisecond to second timescale, forms the basis for advanced neuromorphic functions of the brain, including sound localization, associative learning and working memory [58,59].

At the systems level, the most widely accepted model of human memory and forgetting is the multistore model [60]. As shown in Fig. 1a, new information about input stimuli provided by a "sensor" is first stored in STM and transferred to LTM through a process of rehearsal, where the probability of transfer increases with rehearsal repetitions [47-49]. The similarities between a CuMnAs-based memory device and a biological synapse are schematically illustrated in Figures 1b to 1e.

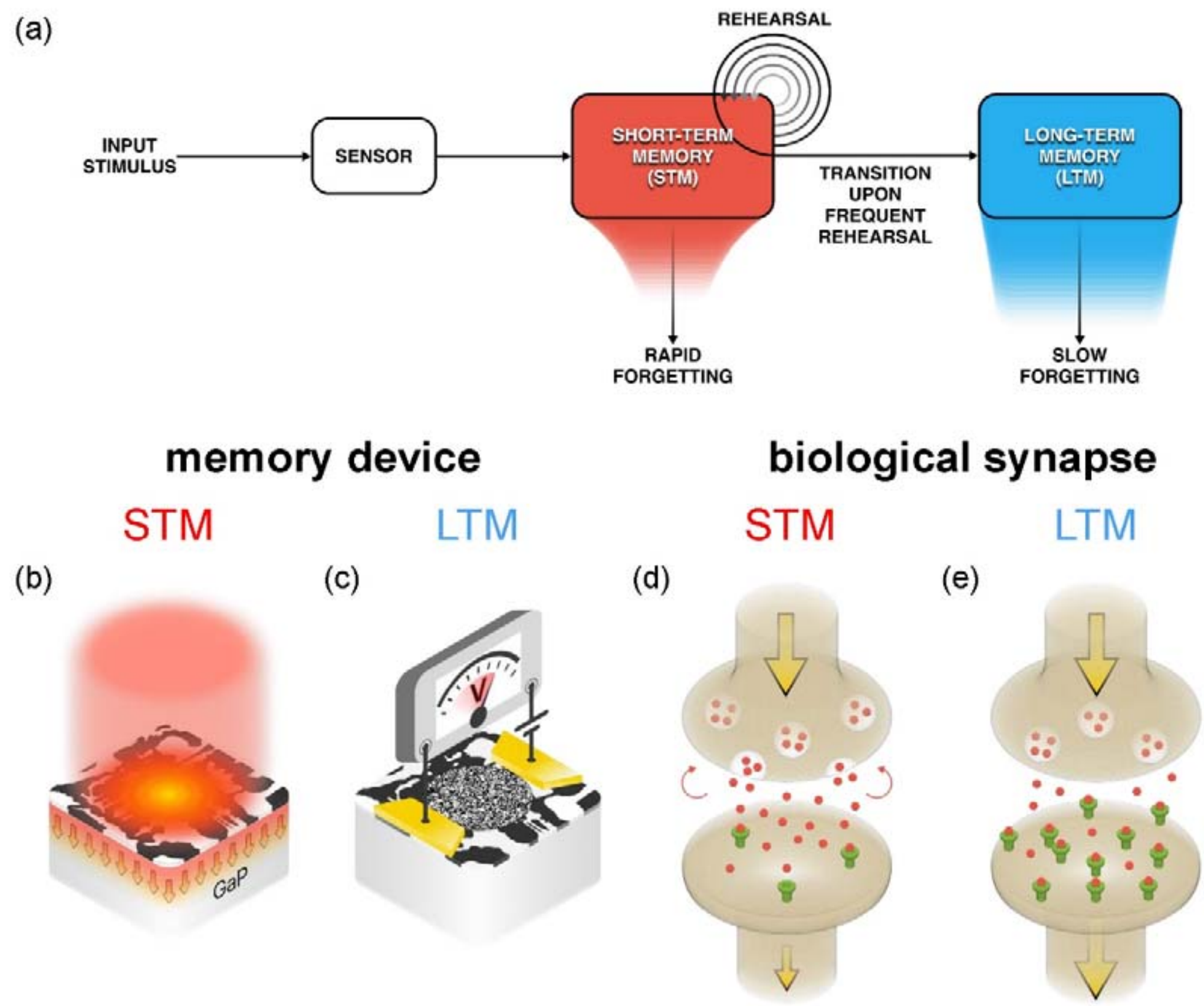

**Figure 1: Illustration of the similarity between a CuMnAs-based memory device and a biological synapse. a**, Multistore psychological model of human memory[60], which illustrates that information from a "sensor" is transferred from short-term memory (STM) to long-term memory (LTM) only after frequent rehearsal. **b** and **c**, Schematic representation of a CuMnAs-based memory device in which the ultrafast change in reflectivity and heat dissipation from the CuMnAs layer (red arrows) after excitation by a single femtosecond laser pulse (red column) play the role of STM and the device resistance increase due to magnetic switching plays the role of LTM[42]. **d,** In a biological synapse, the presynaptic and postsynaptic neurons are separated by a 20 to 40 nm thick gap, called a synaptic cleft. When presynaptic neurons are activated, they release neurotransmitters into the synaptic cleft, which then bind to receptors in a postsynaptic neuron and cause it to become active[58,59]. Since the release of neurotransmitters is initiated by the intracellular calcium concentration ($Ca^{2+}$), the complex kinetics of $Ca^{2+}$ leads to a transient change in the synaptic weight, i.e. STM. **e,** The increase of a number of postsynaptic receptors (shown in green) on longer time scales leads to LTM.

## Experimental setup and samples

In our experiment, we aim at the ultrafast operation of heat-based functionality. Therefore, we use femtosecond laser pulses as excitation stimuli (see Fig. 1c in Ref. 42 and Figs. 1c and 1d in Ref. 46 for time-resolved electrical signals induced in the memory device by the impact of a single femtosecond laser pulse). Standard commercially available femtosecond lasers generate light pulses with repetition rates of only up to ≈ 100 MHz. The mutual time spacing between these pulses is therefore 10 ns or more, which is too long for the planned experiments aimed at compatibility with GHz frequencies of conventional electronics. To address this issue, we have constructed a "multiplier" of laser pulses, which is shown schematically in Fig. 2a.

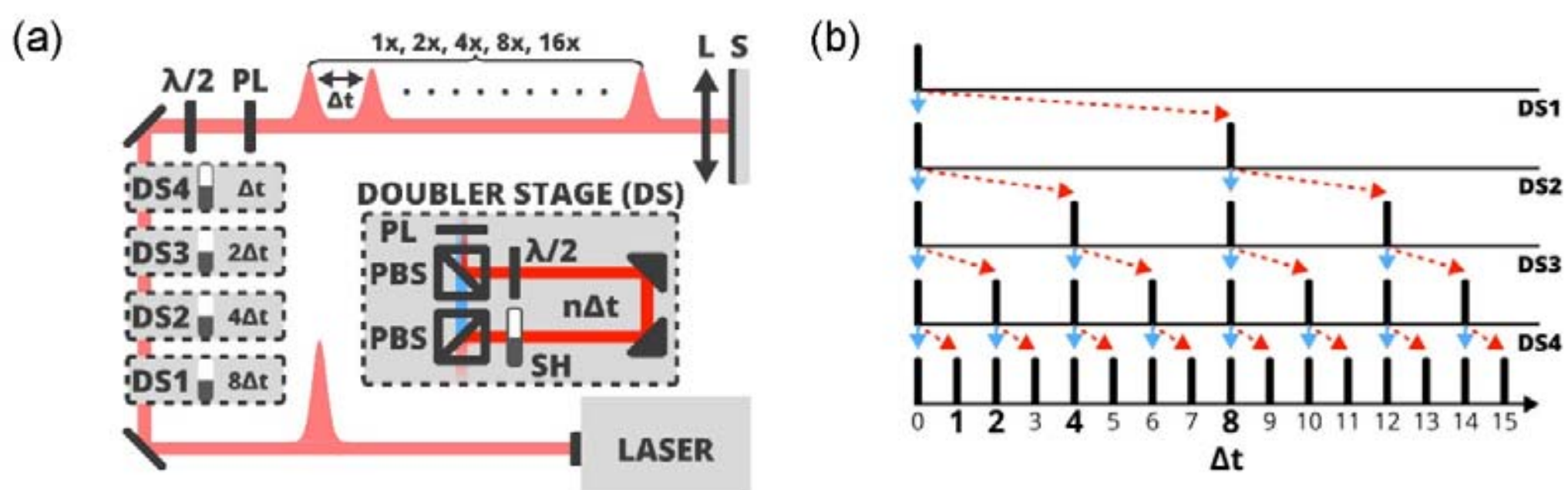

**Figure 2: Schematic representation of the laser-pulse-multiplier**. **a,** A single femtosecond laser pulse is divided into up to 16 equiamplitude and equidistant $\Delta t$-delayed pulses by passing through doubling stages DS1-DS4, each of which causes the indicated time delay, forming the pulse burst focused on the sample (S) by a lens (L). Gray rectangle: A single doubler stage (DS) in which a diagonally polarized laser pulse is split and consequently spatially recombined by a pair of polarization beam splitters (PBS). Here, a delay line introduces a mutual time delay $n\Delta t$ between the transmitted pulse (blue path, horizontal polarization) and the reflected pulse (red path, vertical polarization). A computer-controlled shutter (SH) allows to block the reflected pulse. A half-wave plate or a polarizer (PL) changes the polarization of both pulses to diagonal polarization, allowing a serial concatenation of the stages. **b**, A cartoon showing a single laser pulse splitting across the doubler stages DS1-DS4. Blue solid arrows indicate transmitted pulses and red dashed arrows correspond to reflected pulses, which can be either blocked (state 0) or allowed to pass (state 1) by the shutter SH in each DS.

This setup splits a single 150 fs long laser pulse, generated by our femtosecond laser system with a pulse-on-demand functionality, into its nearly identical copies, resulting in a burst of 1, 2, 4, 8, or 16 laser pulses with a user-selectable equidistant time delay $\Delta t$. A key component in this setup is a pulse doubler stage (DS in Fig. 2a). Here, an incoming laser pulse with diagonal polarization is split into two pulses by a polarization beam splitter (PBS): the transmitted horizontally polarized pulse (blue in Fig. 2a) and the reflected vertically polarized pulse (red). The reflected pulse is either blocked (state 0) by a computer-controlled shutter (SH) or is allowed to pass (state 1) through a delay line to obtain a certain time delay ($n\Delta t$) with respect to the transmitted pulse. Finally, the second PBS recombines the two optical beams. (For more details, see Methods and Supplementary Note 1). Depending on the shutter settings in the four adjacent stages DS1-DS4, bursts of laser pulses are generated with a selected number of femtosecond laser pulses, as shown in Fig. 2b. For example, the generation of 16 pulses in the burst corresponds to shutter setting 1111. Similarly, 8 pulses are generated for 0111, 4 pulses for 0011, 2 pulses for 0001, and a single pulse for 0000. These bursts are used as input stimuli to excite the studied memory devices, whose electrical resistance is measured by the oscilloscope as a function of time (see Fig. 3a).

The studied devices were prepared from epilayers of tetragonal CuMnAs with a thickness of 4 to 60 nm grown by molecular beam epitaxy on a lattice matched GaP substrate [61]. Electron-beam lithography, UV lithography, and wet chemical etching were used to pattern the devices. (For more details on sample preparation see Methods and Ref. 42.) We used a wide range of experimental techniques for the measurements. Dynamics of resistance changes of the devices after their excitation by the laser pulse bursts were measured using a high-performance oscilloscope with a bandwidth of 6 GHz. The heat dissipation from CuMnAs films to the substrate, which strongly depends on the CuMnAs film thickness, was measured by all-optical pump-probe experiments. (For more details, see Methods and Supplementary Note 3.)

**Heat-based memorizing and forgetting in memory devices**

The absorption of a laser pulse in a metallic CuMnAs layer leads to a temporary increase in temperature $\Delta T$ in the illuminated area [62]. If the increased temperature reaches the quench-switching threshold, the device resistance increases [35]. The pulse-induced device resistance increase $\Delta R$ scales linearly with the volume of CuMnAs with a changed resistance (see Supplementary Note S1 in Ref. 42 for a detailed discussion). Interestingly, $\Delta R$ changes smoothly with the laser fluence above the switching threshold, see Fig. 3b. As explained in Supplementary Note 2, this is a consequence of the Gaussian lateral profile of the incident laser pulse and the exponential attenuation during its propagation in the material. Therefore, the switching threshold is only reached in a certain fraction of the illuminated CuMnAs layer, as shown schematically in Fig. 3c. If the time spacing between the excitation pulses ($\Delta t$, see Fig. 3a) is comparable to the time scale of the resistance relaxation towards its equilibrium value, which can be described by a stretched-exponential relaxation on microsecond and longer time scales [35,42], there is a build-up of $\Delta R$ with the number of write pulses. This type of signal buildup corresponds to the magnetic-based LTM functionality, which was studied in detail in Ref. 35 (see Fig. 3b in Ref. 35). In contrast, the main target of this work is the situation where $\Delta t$ is much shorter (i.e., sub-nanosecond) than the magnetic-based dynamics, which corresponds to the heat-based STM functionality.

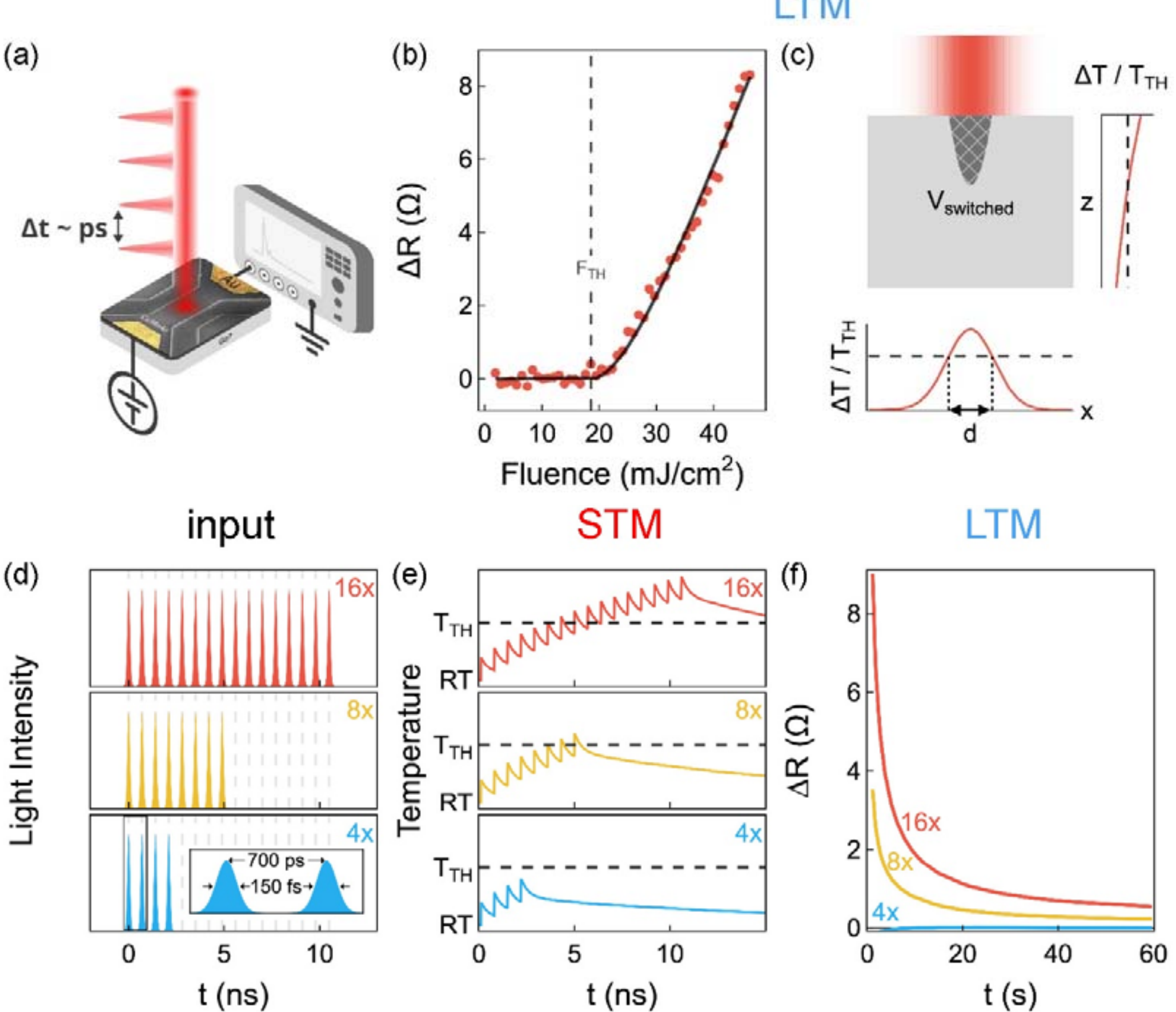

**Figure 3: Heat-based memorizing in a device made from a 20-nm-thick CuMnAs epilayer**. **a,** Schematic of the memory device excited by a burst of femtosecond laser pulses with a mutual time spacing $\Delta t$, leading to a device resistance increase ($\Delta R$) measured with an oscilloscope. **b,** Fluence ($F$) dependence of quench-switching-related $\Delta R$ at 1 µs after the incidence of a single laser pulse, which was deduced from the as-measured $\Delta R$ data as described in Methods (dots). The line is a fit by a theoretical model (see Supplementary Note 2) $\Delta R \sim V_{switched} \sim \ln^2 \left( {}^{F}/_{F_{TH}} \right)$ for $F \geq F_{TH}$, where the single-pulse fluence threshold $F_{TH} \approx 18.5$ mJ/cm$^2$. **c,** Schematic representation of the heating of CuMnAs by a laser. The incident laser pulse with a Gaussian lateral ($x$) profile is absorbed in the CuMnAs epilayer with an exponentially decreasing intensity profile in the light propagation direction ($z$). The volume $V_{switched}$ represents the area in which the temperature increase ($\Delta T$) of CuMnAs induced by the laser pulse exceeds the threshold temperature ($T_{TH}$). Within this volume, the material resistivity increases, leading to long-term memory (LTM) functionality. **d-f,** Experimental demonstration of a psychological model of memorizing and forgetting (depicted in Fig. 1a) in the memory device. **d,** Laser pulse bursts are used as input stimuli and the absorption in CuMnAs serves as a sensor; note that the pulse widths are greatly exaggerated with respect to their time spacing for graphical clarity (see inset in **d**). **e,** Schematic representation of a transient rise in effective CuMnAs temperature after absorption of the number of laser pulses shown; $T_{TH}$ is exceeded only for 8 and 16 pulses at room temperature (RT). **f,** Measured resistivity increase $\Delta R$ shows that within the used 10-ns-long memorizing period, at least 8 rehearsals are required for information to be transferred from STM to LTM using laser pulses with a fluence of 0.5 $F_{TH}$.

The experimental demonstration of the applicability of the psychological model of memorizing and forgetting (Fig. 1a) to heat-based information encoding is shown in Figures 3d-3f. We use femtosecond laser pulses with a selectable number of pulses in each burst as input stimuli (Fig. 3d), while the absorption in CuMnAs serves as a "sensor". As the number of laser pulses (i.e., the rehearsal number) increases, the transiently accumulated temperature in the CuMnAs film increases, as shown in Fig. 3e. When the switching threshold is reached, nanoscale magnetic textures in CuMnAs are formed which, in turn, lead to an increase in device resistance that is electrically detectable on extended times (Fig. 3f). A further increase in the number of laser pulses increases the volume of the material in which the switching threshold is exceeded and thus the measured device resistance increases. As described in detail in Ref. 42, heat dissipation in CuMnAs films after absorption of a femtosecond laser pulse takes place on several different time scales. The dominant processes, which are exploited in this paper, can be approximated by a sum of three exponential decay functions with characteristic time constants $\tau_i$ (and amplitudes $A_i$): The time constant $\tau_1$ (with amplitude $A_1 \sim 0.67$), which describes the heat transfer from the CuMnAs film to the GaP substrate, strongly depends on the film thickness, reaching ~ 400 and 1 300 ps in the 20- and 50-nm films, respectively. (The full thickness dependence of $\tau_1$, determined from the measured dynamics of transient absorption illustrated in Supplementary Fig. S3, is shown in the inset of Fig. 4a.) On the other hand, time constants $\tau_2 \sim 10$ ns ($A_2 \sim 0.25$) and $\tau_3 \sim 100$ ns ($A_3 \sim 0.08$), which were attributed to the heat transfer from the GaP substrate to the sample holder and environment [42], do not depend on the film thickness [42]. In Fig. 4a, we compare the cooling dynamics in 20-nm-thick and 50-nm-thick CuMnAs epilayers on (sub)nanosecond time scales. Due to the film-thickness-dependent time constant $\tau_1$, these dynamics strongly depends on the film thickness up to ~ 4 ns . To demonstrate the tunability of the STM-dynamics by the CuMnAs film thickness, without affecting the LTM-dynamics, we used a 700 ps time spacing between the laser pulses in the experiments shown in Figs. 3 and 4 (see Supplementary Note 4 and Supplementary Fig. S4 for an explanation of why 700 ps is the optimal time spacing for achieving this goal in CuMnAs). As can be seen in Fig. 4b, the required switching threshold temperature is exceeded at 8 incident pulses for both film thicknesses. However, as the threshold is only just reached for the 20 nm film, the resulting long-term resistance increase is much larger for the device made from the 50 nm film (but the LTM-dynamics are very similar in both cases), see Fig. 4c.

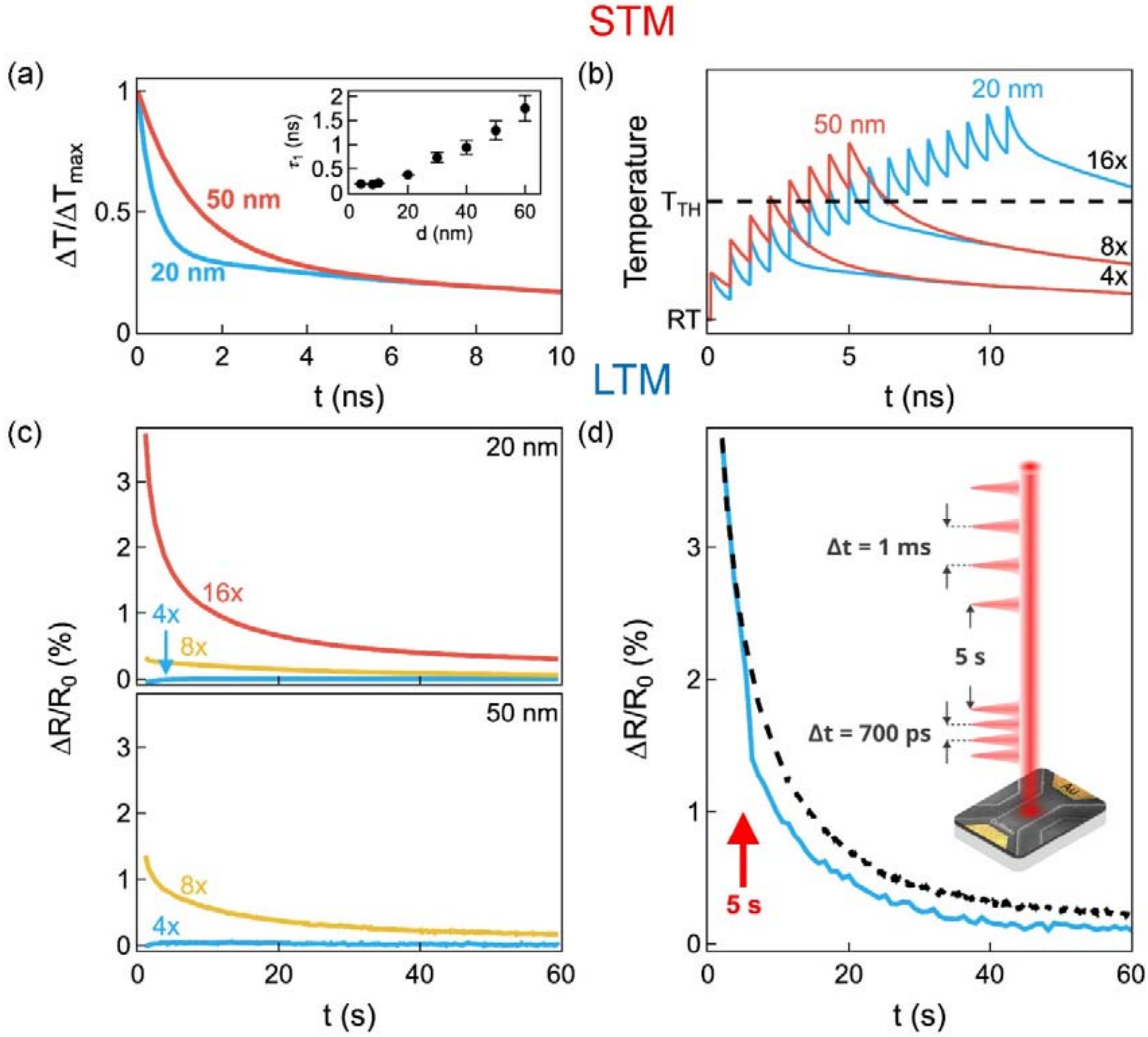

**Figure 4: Heat-based control of memorizing and forgetting. a**, Cooling dynamics in CuMnAs films with thicknesses of 20 and 50 nm, as approximated by a sum of 3 exponential decay functions, see text for details. Inset: Dependence of the time constant $\tau_1$ on the epilayer thickness; the error bars depict the experimental uncertainty of the $\tau_1$ determination from the pump-probe experiment (see Supplementary Note 3 for details about this experiment). **b**, Schematic representation of the transient temperature evolution in CuMnAs films with thicknesses of 20 and 50 nm, illustrating the STM dynamics, after excitation by laser pulse bursts with a depicted number of femtosecond laser pulses with a time spacing of 700 ps (see Fig. 3d). At room temperature (RT), the temperature threshold ($T_{TH}$) required to achieve the switching is reached for 8 pulses for both film thicknesses. However, as the threshold is only barely exceeded in the 20-nm film, the resulting long-term relative resistive increase is much larger in the device fabricated from the 50-nm film, as shown in **c** for pulses with a fluence of 0.4 $F_{TH}$. **d**, Demonstration of laser-induced erasing of information stored in LTM. The dashed curve shows the resistivity decay (i.e., a "forgetting" dynamic) measured after a 10 ns long memorizing period, when 16 rehearsals by pulses with a fluence of 0.4 $F_{TH}$ and a time spacing of 700 ps were used in a device fabricated from a 20-nm-thick film. The solid curve shows that forgetting can be significantly accelerated if at a given time point (at 5 s in **d**, as indicated by a vertical arrow) laser pulses with the same fluence but with a much larger mutual time spacing of 1 ms start to be incident on the memory device. The results for other time spacing between the erasing pulses are shown in Supplementary Fig. S5.

Finally, we show that it is also possible to rapidly reset the device from the metastable high-resistive state to the low-resistive ground state using laser pulses. In Fig. 4d, the dashed curve shows the forgetting in the LTM after a 10 ns long memorizing period, in which 16 rehearsals by laser pulses time-spaced by 700 ps were performed to achieve information transfer from the STM to the LTM. The solid line illustrates that LTM forgetting can be significantly accelerated by laser pulses. In this case, laser pulses with the same intensity but with a much longer mutual time spacing, then those used for the rehearsals (1 ms vs. 700 ps), started to be incident on the device 5 s after the information was written, as shown by a vertical arrow in Fig. 4d. As discussed in detail in Supplementary Note 5, this is a consequence of the strongly temperature-dependent relaxation rate from the quench-switched states (see Figs. 4a and 4b in Ref. 35). Importantly, this LTM reset can be applied multiple times without device degradation (see Fig. 5c in Ref. 35). When the device temperature returns to its original level, the memory properties are fully restored.

## Discussion and conclusion

Ultrafast control of magnetic order by femtosecond laser pulses has attracted attention since its discovery in 1996, when demagnetization of a thin Ni film in less than 1 ps was demonstrated [63]. In particular, all-optical magnetization switching [64-66] is considered the least-dissipative and fastest method for magnetic writing in memory applications [67,68]. Laser pulses can interact with magnetic systems through various mechanisms [63,67,69], with thermal effects being by far the most common in metallic materials.

STM to LTM transfer is a topic intensively studied in bio-inspired synaptic devices, but only on millisecond and longer time scales [47-49,70-73]. Our experimental results show that very similar phenomena and operating principles can also be found in analog CuMnAs-based memory devices, but the relevant (picosecond) timescales are several orders of magnitude faster. Interestingly, the wavelength-dependent transient change of the CuMnAs reflectivity, which can be either positive or negative [62], allows control over what fraction of the incident laser pulse energy is transferred to the generated heat pulse that induces the actual CuMnAs switching [42]. Since this effect is controlled by the electron-phonon relaxation with a characteristic time constant of 2.4 ± 0.3 ps in CuMnAs [62], the STM properties can be altered on a picosecond time scale [42]. This behavior resembles the short-term plasticity of synaptic devices, but on a much shorter time scale – see Fig. 3 in Ref. 58, where the current status in the field of short-term artificial synaptic devices is reviewed.

In this study, we used heat pulses generated by absorption of femtosecond laser pulses in the metallic CuMnAs layer. Due to the ratio between the diameter of the excitation spot of 20 μm and the thicknesses of the CuMnAs films of a few tens of nm, the one-dimensional heat diffusion from the CuMnAs film to the GaP substrate dominates the dynamics of heat dissipation (see Appendix C in Ref. 62 for the modeling of heat dissipation). Consequently, these dynamics can be controlled by the film thickness (cf. inset in Fig. 4a and Supplementary Fig. S3) and approaches ~ 100 ps for the sub-10-nm-thick layers, making the heat-based STM fully frequency compatible with the GHz frequencies of conventional electronics. If required, the dissipation dynamics can be further accelerated by the presence of a heat sink, similarly as in the HAMR technology in HDD [12].

At present, it is difficult to imagine the use of CuMnAs-based devices in neuromorphic computing systems, where the output of one device should serve as the input for the subsequent device. On the other hand, these memory devices can serve as heat-to-charge conversion units, where thermal logical values in heat circuits are converted to electrical ones, a necessary step to achieve compatibility of thermal logic with standard electronics. Moreover, they simultaneously provide the heat-based logic-in-memory functionality.

Finally, we address the prospects for the miniaturization of CuMnAs devices. The current method of device fabrication from thin films enables the preparation of devices with characteristic dimensions of down to ~ 100 nm (see Fig. 4 in Ref. 46), but the process can be further optimized to achieve smaller devices. The eventual miniaturization of the devices is important not only to achieve their higher density, but also from the point of view of energy efficiency – the energy required for switching is considerably lower in smaller devices. Considering the nanometer-sized domains [74] and the presence of atomically-sharp domain walls [75] in CuMnAs antiferromagnet, it is reasonable to expect that the device size could be potentially scaled down to ~ nm dimensions, which leads to the switching energy of ~ 3.5 fJ in CuMnAs [42]. This energy is significantly lower than the current energy cost limits of ~ 150 fJ per floating point operation in modern CMOS microprocessors [76] and is comparable to the ~ 1-100 fJ per synaptic event in biological systems [77] and ≤ 100 fJ switching cost in the most energy-efficient magnetic memories [78]. As for the scaling of the laser beam size, this is usually restricted by the diffraction limit in standard (far field) optical setups. Smaller laser spot sizes in the order of ~ 10 nm can be provided in near field setups for scanning probe microscopy [79], where light is focused by a metal tip. On-chip integration of CuMnAs-based memories in conceivable in waveguide-based photonic circuits in which sub-wavelength resolution can be easily achieved, as has been experimentally demonstrated in optically-addressed memories based on phase-change materials [45,80].

## Methods

### Samples

Tetragonal CuMnAs epitaxial layers were grown on lattice-matched GaP substrate at temperatures around 200 °C by molecular beam epitaxy. The films were capped with a 3 nm thick Al layer, which oxidized rapidly after removal from vacuum, preventing oxidation of the CuMnAs. Four-terminal devices with typical dimensions of 20 × 20 μm$^2$, optimized for four-point resistance measurements, were lithographically patterned from 20-nm-thick epilayers (with a sheet resistance of 50 Ω) by electron-beam lithography, UV lithography, and wet chemical etching. The aluminum cap was removed by an alkaline developer during the resist development phase and CuMnAs layer was etched by wet etching in a solution of $H_3PO_4$.

### Pump-probe experiment

To investigate the heat dissipation dynamics in CuMnAs films of different thicknesses, we performed a degenerate pump-probe experiment using a femtosecond Ti:Sapphire oscillator (Mai Tai, Spectra Physics) generating ≈ 150 fs laser pulses at 820 nm with a repetition rate of 80 MHz. The pump and probe pulses were focused on the same spot (with a Gaussian diameter of ≈ 35 μm) on the sample at nearly normal incidence, see Fig. 1a in Ref. 81 for the experimental setup used. The laser fluence of the pump pulses was approximately 3 mJ/cm² (i.e. well below the CuMnAs switching threshold), while the probe pulses were at least 50 times weaker. The samples were placed in a closed-cycle helium cryostat (ARS) at temperature of 15 K. We measured the pump-induced differential transmittance $d\mathcal{T}/\mathcal{T} = (\mathcal{T}_E - \mathcal{T})/\mathcal{T}$, where $\mathcal{T}_E$ and $\mathcal{T}$ are transmittances with and without the pump pulse, respectively.

### Device resistance changes after excitation by bursts of femtosecond laser pulses

A Yb-based femtosecond laser system (Pharos, Light Conversion) was used as the light source for the optical excitation of the investigated devices. This laser can generate a single laser pulse with a 150 fs pulse duration and a central wavelength of 1030 nm using the pulse-on-demand function of the control software. The single laser pulse was split into $N$ identical copies, separated by equidistant time delays $\Delta t$ = 700 ps, using a pulse-multiplier experimental setup, as shown in Fig. 2 and described in detail in Supplementary Note 1. In our experiment, each laser pulse burst contained $N$ = 1, 2, 4, 8, or 16 laser pulses with a computer-controlled laser fluence. The pulses were focused on a single spot on the sample, which had a Gaussian intensity profile with a diameter of ≈ 20 μm, to excite the studied devices. A four-point measurement configuration (using a Keithley 6500 multimeter and a Keithley 2400 sourcemeter with 100 μA probing current) was used to record the laser-induced change in device resistance in a real time every 0.1 s for 60 s after the burst arrival. For a given laser pulse fluence and a selected number of $N$ pulses in a burst, three acquisitions were typically measured and averaged to improve data quality. Note that the reproducibility of the laser-induced device resistance change was studied in detail in Ref. 46, where standard deviations from 0.10 to 0.12 of the normalized resistivity changes were obtained (see Fig. 2d in Ref. 46). All measurements were performed at room temperature.

The impact of intensive laser pulse on CuMnAs-based memory device leads to its resistance increase $\Delta R$ due to a heat-dependent CuMnAs resistance (see Supplementary Fig. S4a in Ref. 42) and quench-switching, which can be experimentally separated as described in detail in Ref. 42 (see Figure 2 therein). For clarity, in Fig. 3b we show only the quench-switching-related $\Delta R$ signal, which was obtained from the as-measured $\Delta R$ data by subtracting the linear heating contribution[42]. In all other figures, we show the as-measured $\Delta R$ data (sometimes plotted as a relative resistance change $\Delta R / R_0$, where $R_0$ is the equilibrium device resistance).

**Abbreviations**

| | |
|---|---|
| IT | information technology |
| HAMR | heat-assisted magnetic recording |
| HDD | hard-disk drive |
| AF | antiferromagnets |
| AMR | anisotropic magnetoresistance |
| STM | short-term memory |
| LTM | long-term memory |
| DS | doubler stage |
| PBS | polarization beam splitter |
| SH | shutter |
| S | sample |
| L | lens |
| RT | room temperature |

## Supplementary Information

The online version contains supplementary material available at [to be provided]. Supplemental materials consist of Supplementary Note 1. Pulse-multiplier experimental setup (containing Figure S1), Supplementary Note 2. Volume of the switched CuMnAs (containing Figure S2), Supplementary Note 3. Measurement of heat dissipation speed from CuMnAs films to GaP substrate (containing Figure S3), Supplementary Note 4. Dependence of STM dynamics on pulse spacing (containing Figure S4), and Supplementary Note 5. Heat-based control of LTM erasing (containing Figures S5 and S6), and Supplementary references.

### Authors' contributions

P.N., M.S., K.O., J.Z., L.N., and T.J. planned the experiments, V.N., F.K., A.F., and J.Z. prepared the samples, M.S., A.F. and J.Z. prepared the software for the data acquisition, M.S., A.F., and P.K. performed the experiments, T.O. provided theoretical support, P.N., M.S. and T.J. wrote the manuscript with contributions from all authors.

### Funding

This work was supported by TERAFIT project No. CZ.02.01.01/00/22_008/0004594 funded by Ministry of Education Youth and Sports of the Czech Republic (MEYS CR), programme Johannes Amos Comenius (OP JAK), call Excellent Research. The authors acknowledge funding by the Czech Science Foundation (grant no. 21–28876J), by ERC Advanced Grant no. 101095925, by the Grant Agency of the Charles University (grants no. 166123 and SVV–2024–260720), by CzechNanoLab Research Infrastructure supported by MEYS CR (LM2023051), and by MEYS CR project LNSM-LNSpin.

This version of the article has been accepted for publication, after peer review but is not the Version of Record and does not reflect post-acceptance improvements, or any corrections. The Version of Record is available online at: http://dx.doi.org/10.1186/s43074-025-00207-1.

### Data availability

Data reported in this paper are available in the Zenodo repository [10.5281/zenodo.17357690] and are publicly available as of the date of publication.

## Declarations

### Competing interests

The authors declare no competing interests.

# Sub-nanosecond heat-based logic, writing and reset in an antiferromagnetic magnetoresistive memory: Supplementary information

M. Surýnek[1], A. Farkaš[2,1], J. Zubáč[2,1], P. Kubaščík[1], K. Olejník[2], F. Křížek[2], L. Nádvorník[1], T. Ostatnický[1], R.P. Campion[3], V. Novák[2], T. Jungwirth[2,3], and P. Němec[1,*]

[1]Faculty of Mathematics and Physics, Charles University, Ke Karlovu 3, 121 16 Prague 2, Czech Republic

[2]Institute of Physics ASCR, v.v.i., Cukrovarnická 10, 162 53 Prague 6, Czech Republic

[3]School of Physics and Astronomy, University of Nottingham, Nottingham NG7 2RD, United Kingdom

[*] Electronic mail: petr.nemec@matfyz.cuni.cz

**Supplementary Note 1. Pulse-multiplier experimental setup**

As described in the main paper, to be compatible with GHz frequencies of conventional electronics, we aimed at experiments with bursts of laser pulses with a mutual time spacing in sub-nanosecond time range. However, standard commercially available femtosecond lasers generate light pulses with repetition rates of only up to ≈ 100 MHz, where the time spacing between these pulses is 10 ns or more. Therefore, we had to construct the pulse-multiplier experimental setup, described in this text and depicted in Fig. S1a, that splits a single 150 fs long laser pulse, generated by our femtosecond laser system with a pulse-on-demand functionality, into its nearly identical copies. A key component in this setup is a doubler stage (DS in Fig. S1a). Here, an incoming laser pulse with a diagonal polarization is split by a polarization beam splitter (PBS) into two pulses, the transmitted horizontally-polarized pulse (blue in Fig. S1a) and the reflected vertically-polarized time-delayed pulse (red in Fig. S1a), which are spatially recombined by the second PBS. Seemingly, this might look as a quite straightforward task. In reality, however, this is a rather challenging achievement due to the non-ideal properties of commercially available PBS. In particular, the plate-shaped PBS distort the spatial profile of transmitted/reflected beams considerably due to the not-normal angle of incidence on their surface, which does not enable the needed perfect spatial overlap of the spatially recombined pulses after the second PBS without performing their Fourier-based spatial filtration. On the other hand, the cube-shaped PBS, where the beam spatial distortion is absent due to the normal angle of incidence, do not have an ideal polarization properties because they are optimized only with respect to the polarization of transmitted light while the polarization-quality of reflected light is much lower and, which is even worse, the polarization plane of these to beams is *not* mutually perpendicular (typically, the deviation from the "expected" angle between the corresponding polarization planes of 90 deg is several degrees). Consequently, after a serial concatenation of several doubler stages this leads to considerable intensity variations between individual pulses within the generated laser pulse train. In our experimental setup, we opted for a cube-shaped PBS where we rotated additionally the reflected light polarization plane by a half-wave plate (see Fig. S1a), to have it exactly at 90 deg with respect to the polarization plane of the transmitted light. After the 2$^{nd}$ PBS, a half-wave plate is used to change the polarization of both pulses to a diagonal one, which allows for their further splitting in the following DS. Nevertheless, as the polarization of beam reflected at PBS acquires also some ellipticity, it is advantageous to replace the half-wave plate by a diagonally oriented polarizer at least after a pair of DS, which purifies the linear polarization of all splitted laser pulses and ensures the correct functionality of the following DS. By this approach, we were able to generate 16 laser pulses with very similar intensities that were co-propagating for several meters with nearly identical spatial profiles and having the selected mutual time spacing between them (see Fig. S1c).

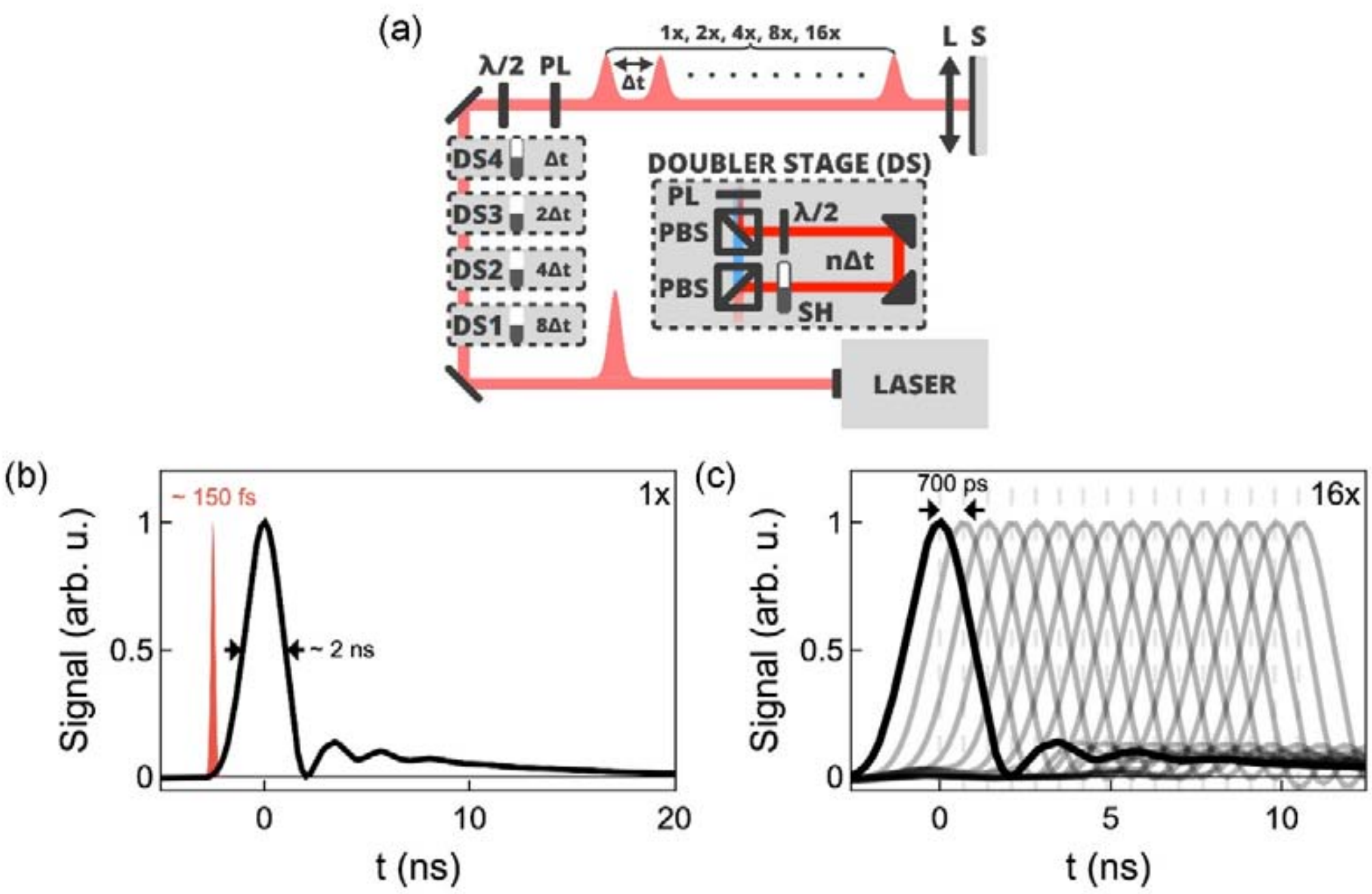

**Supplementary Figure S1. Generation of a burst of laser pulses from a single laser pulse. a,** Schematics of the corresponding experimental setup. A laser pulse is divided into up to 16 equiamplitude and equidistant $\Delta t$-delayed pulses by passing through doubler stages DS1-DS4, each of them inducing the indicated time delay, which form the pulse burst incident on the sample. Grey rectangle: An individual doubler stage (DS) where a diagonally polarized laser pulse is split and consequently spatially recombined by a pair of polarization beam splitters (PBS). Here, a delay line introduces a mutual time delay n$\Delta t$ between the transmitted (blue path, horizontal polarization) and reflected pulse (red path, vertical polarization). A computer-controlled shutter (SH) allows to block the reflected pulse. A half-wave plate ($\lambda/2$) or a polarizer (PL) changes the polarization of both pulses to a diagonal one, allowing a serial concatenation of DS. **b**, Time-resolved signal measured by an avalanche photodiode (black) after impact of one 150-fs-long laser pulse (red). **c**, Signals measured for each of the 16 individual laser pulses from the burst, which are mutually time delayed for $\Delta t$ = 700 ps.

**Supplementary Note 2. Volume of the switched CuMnAs**

The laser pulses were focused by lens on the device structure, producing a Gaussian intensity profile

$$I(r) = I_0 \cdot e^{-\frac{r^2}{w^2}} \qquad \text{(S2.1)}$$

where $r$ is the radial distance from the center of the Gaussian profile. The experimentally achieved beam size, characterized by a full-width $2w \approx 20$ µm, was determined using the knife-edge method. Due to the absorption in CuMnAs epilayer, the pulse intensity decays exponentially with an absorption coefficient of $\alpha = 3.1 \cdot 10^5$ cm$^{-1}$ [S1]. The resulting energy distribution of the laser light (expressed as energy density per unit volume) within the epilayer is described by

$$U(r,z) = U_0 \cdot e^{-\frac{r^2}{w^2}} \cdot e^{-\alpha z}, \qquad \text{(S2.2)}$$

where $z$ denotes distance in the epilayer along the laser propagation direction, and $U_0$ is the amplitude of the energy density. The measured device resistance change $\Delta R$ is linearly proportional to the volume of the region in which the energy density in CuMnAs surpasses a threshold value, $U(r, z) \geq U_{TH}$. This volume is given by

$$V_{\text{switched}} = \frac{\pi w^2}{2\alpha} \ln^2(\theta)\,;\ \theta \geq 1, \qquad \text{(S2.3)}$$

where $\theta$, defined as $\theta = U_0 / U_{TH}$, measures the extent to which the threshold energy is exceeded. This quantity can also be represented in terms of the incident laser fluence $F$ (energy density per unit area) as $\theta \approx F / F_{TH}$, where $F_{TH}$ is the threshold fluence. If $\theta < 1$, the threshold condition for achieving the switching is not fulfilled, implying $V_{\text{switched}} = 0$. This switched volume can be envisaged as a cap-like shape (see Fig. S2) with a circular base of diameter $d$:

$$d = 2w\sqrt{\ln(\theta)};\quad \theta \geq 1. \qquad \text{(S2.4)}$$

Fig. S2b demonstrates how the diameter $d$ expands as the switching threshold is exceeded. Despite the relatively large spot size of the incident laser pulse, a considerably smaller region can be switched.

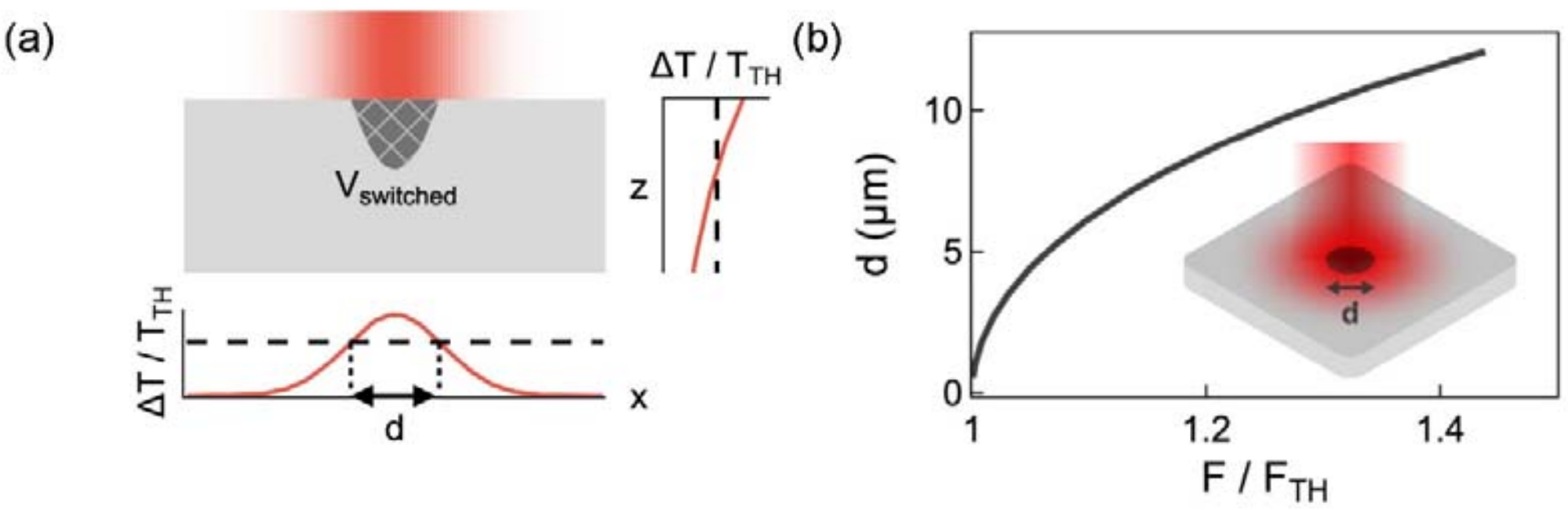

**Supplementary Figure S2. Volume of CuMnAs with a resistance increased due to the quench-switching. a,** The incident laser pulse with a Gaussian lateral ($x$) profile is absorbed in the CuMnAs epilayer with an exponentially decaying intensity profile in the $z$-direction. The volume $V_{\mathrm{switched}}$ represents the region where the laser-pulse-induced temperature increase $\Delta T$ of CuMnAs exceeds the threshold temperature $T_{\mathrm{TH}}$. This volume can be visualized as a cap-like structure with a circular base of diameter $d$. **b,** The dependence of $d$ on fluence $F$, expressed relative to the value of the threshold fluence $F_{\mathrm{TH}}$.

**Supplementary Note 3. Measurement of heat dissipation speed from CuMnAs films to GaP substrate**

The photoexcitation of a metal by an intense femtosecond laser pulse excites the electron distribution out of equilibrium on a time scale much shorter than the electron–phonon interaction time. The resulting non-thermal population of electrons thermalizes rapidly by electron–electron scattering processes. Consequently, a thermalized electron system, which can be described by a Fermi distribution with an electron temperature, is formed within ≈ 100 fs after the impact of the pump pulse. On a picosecond time scale, the excess energy is dissipated from the electron system to the lattice by electron–phonon scattering processes, which leads to an increase in the lattice temperature. Finally, heat diffusion dissipates the excess energy and the metal returns to the equilibrium state. Importantly, all the above effects lead to a change in optical properties and, therefore, the corresponding characteristic time constants can be evaluated from the measured optical transient signals [S1]. In Fig. S3 we show results of degenerate pump–probe experiment in in CuMnAs films of different thicknesses where a time evolution of pump-induced transmission change was measured by probe pulses of the same wavelength as that of the pump pulse. The lines are fits by a mono-exponential decay function with depicted characteristic time constants, which describe the heat dissipation from the CuMnAs epilayer to the GaP substrate. Clearly, the heat dissipation is much faster in thinner films.

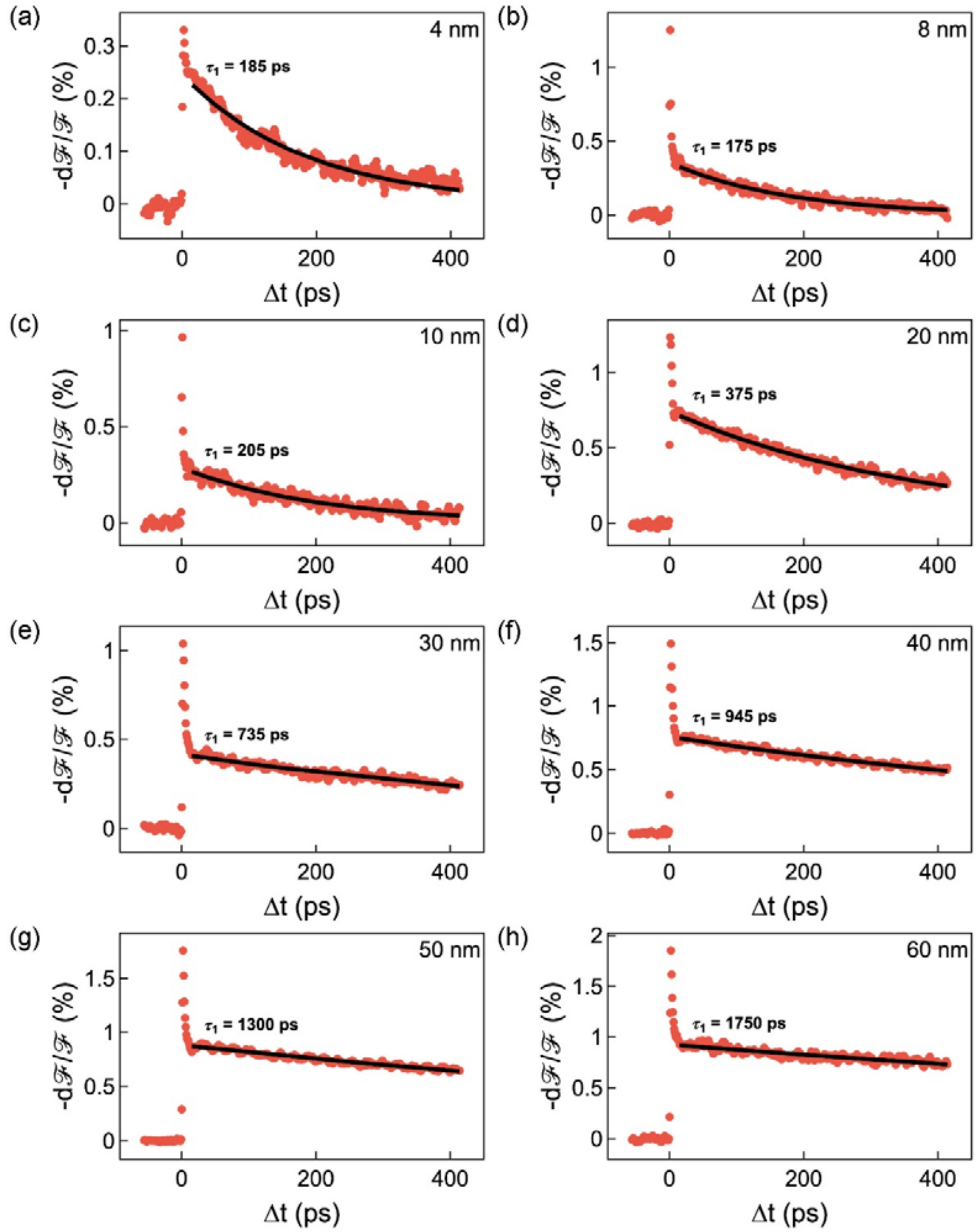

**Supplementary Figure S3. Measurement of heat dissipation in CuMnAs films of different thicknesses.** Laser-pulse-induced dynamics of transient decrease of differential transmission $d\mathcal{F}/\mathcal{F}$ measured at 15 K by optical pump-probe experiment in CuMnAs epilayers with indicated thicknesses **a-f**, using pump and probe pulses with a wavelength of 820 nm (dots) and pump fluence of 3 mJ.cm$^{-2}$. Lines are fits by a mono-exponential decay function with a characteristic time constant $\tau_1$.

**Supplementary Note 4. Dependence of STM dynamics on pulse spacing**

The heat-based short-term memory (STM) functionality in our memory device is governed by an interplay between heat accumulation in CuMnAs film from successive femtosecond laser pulses and heat dissipation into the substrate and environment. Following the procedure described in detail in Ref. S3, we model the heat dynamics after absorption of a laser pulse by a tri-exponential relaxation comprising a film-thickness-dependent heat dissipation from the film to the substrate, which is characterized by a time constant $\tau_1$, and two slower terms. These two terms, with time constants $\tau_2$ and $\tau_3$, are associated with a heat transfer from the substrate to the sample holder and environment, which do not depend on the film thickness. For the CuMnAs films used in our experiment, $\tau_1 \approx 375$ ps and 1 300 ps for the 20-nm-thick and 50-nm-thick films, respectively, (see Fig. S3) with a relative amplitude $A_1 = 0.67$. The slower components are ~10 ns and ~100 ns (see Fig. 2c in Ref. S3) with relative amplitudes $A_2 = 0.25$ and $A_3 = 0.08$, respectively. On the sub-microsecond timescales relevant to STM, all eventual slower dissipation channels to environment can be disregarded. However, this residual heat can be used for the long-term memory erasing, as discussed in Supplementary Note 5.

For a device excitation by multiple laser pulses with a fixed time spacing $\Delta t$, we assume a linear superposition of single-pulse thermal responses. The quench switching in CuMnAs occurs when the cumulative thermal response after $n$ pulses $S_n(\Delta t)$ exceeds the threshold temperature $T_{TH}$. In our model, the relative value of $T_{TH}$ and single-pulse heating amplitude $A$ were calibrated using the 20-nm film experimental data shown in Fig. 3f of the main paper: for $\Delta t$ = 700 ps, the switching onset occurs between 4 and 8 pulses. In Figs. S4a – S4c we show how the balance between the heat accumulation and dissipation is controlled by the mutual pulse spacing $\Delta t$ for 20 nm and 50 nm CuMnAs films. When $\Delta t$ = 100 ps (i.e. shorter than $\tau_1$), 3 pulses are required to reach the threshold temperature for both film thicknesses, see Fig. S4a. When $\Delta t$ = 5 000 ps (i.e. much longer than $\tau_1$), 21 pulses are needed to achieve switching, again independently of the film thickness, see Fig. S4c. Only when $\Delta t$ is comparable to $\tau_1$, e.g. 700 ps as in Fig. S4b, the intended sensitivity of the number of pulses required for reaching the switching on the CuMnAs film-thickness is achieved. To put this analysis on a more quantitative basis, we plot in Fig. S4d the difference between the number of needed optical pulses ($\Delta NOP$) in 50-nm and 20-nm-thick films. However, as this number is strongly sensitive to the exact choice of $T_{TH}$ and $A$ in our model calculations, this analysis was repeated for 1 000 uniformly sampled threshold temperatures within interval $[T_{TH} - A/2, T_{TH} + A/2]$, where $A$ is the single-pulse induced heating.

The obtained values of Δ*NOP* are shown as points in Fig. S4d. From these values the average value ⟨Δ*NOP*⟩ can be computed, which is depicted as a line in Fig. S4d. This dependence reaches a maximum at $\Delta t \approx 700$ ps, which is the time delay between pulses used to obtain the experimental results shown in Figs. 3 and 4 of the main paper.

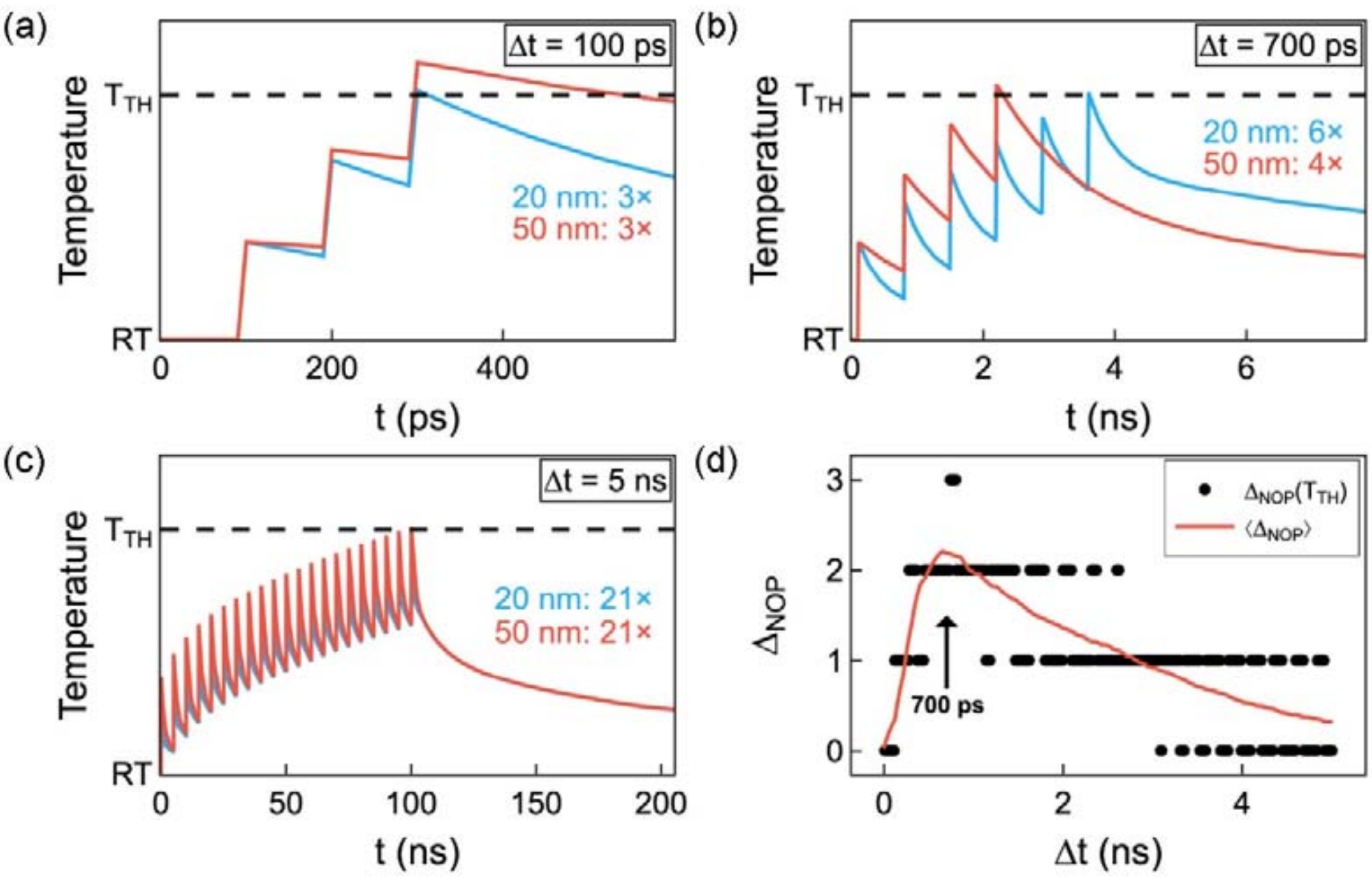

**Supplementary Figure S4. Dependence of STM functionality on the pulse spacing between successive femtosecond laser pulses. a**–**c**, Modeled cumulative temperature for 20-nm (blue) and 50-nm-thick (red) CuMnAs films exposed to bursts of laser pulses with different time spacing $\Delta t$: **a,** $\Delta t = 100$ ps, **b,** 700 ps, and **c,** 5 ns. Depicted numbers indicate how many laser pulses are needed for exceeding the threshold temperature $T_{TH}$ for the corresponding film thickness. **d,** Difference between the number of needed optical pulses (NOP) for achieving the switching in 50-nm and 20-nm-thick films as a function of $\Delta t$ (points), see text. Line is a threshold-averaged dependence that has a maximum around $\Delta t = 700$ ps, which was used to achieve the experimental results shown in Figs. 3 and 4 in the main paper.

**Supplementary Note 5. Heat-based control of LTM erasing**

As discussed in detail in the main paper, the incident laser pulses can lead not only to information memorizing (i.e., the increase of the device resistivity) but also to an acceleration of information forgetting, depending solely on the time spacing between the illuminating pulses. In this experiment, we used a 10 ns long memorizing period, when 16 rehearsals by laser pulses time-spaced by 700 ps were performed to achieve information transfer from the STM to the LTM. In Fig. 4d of the main paper, we illustrated that LTM forgetting can be significantly accelerated by laser pulses with the same intensity but with a time spacing of 1 ms. The reason why laser pulses with a time spacing of 700 ps cause the resistivity increase while the same pulses spaced by 1 ms cause the resistivity decrease lies in the non-commensurate time scales of the heat-related and quench-switching-related dynamics. If the pulse spacing is shorter than the time constants describing the main components of the heat-based dynamics of STM (i.e. (sub)nanosecond), the device excitation by a train of laser pulses leads to a temperature buildup, which induces the quench-switching in CuMnAs (see Fig. 3 in the main paper and Fig. S4). If the pulse spacing is longer (e.g., 1 ms), the accumulated temperature does not overpass the threshold temperature for the quench-switching but it leads to an acceleration of the strongly temperature-dependent relaxation rate from the high-resistive quench-switched states (see Figs. 4a and 4b in Ref. S2), as shown in Fig. 4d in the main paper. For even longer pulse spacing (above ~ 10 ms), also this effect disappears and the additional pulses have virtually no effect, as illustrated in Fig. S5a. To complete the picture, in Fig. S5b we show the results for the pulse spacing of 0.1 ms. Also in this case the accumulated temperature is not high enough to overpass the switching threshold and, consequently, the resistivity decrease is observed due to the laser-induced acceleration of the relaxation rate. Nevertheless, as the (quasi)equilibrium resistance of the metallic CuMnAs increases with the temperature (see Supplementary Fig. S4a in Ref. S3), the device resistance does not relax to the original equilibrium value but to a value corresponding to slightly heated CuMnAs.

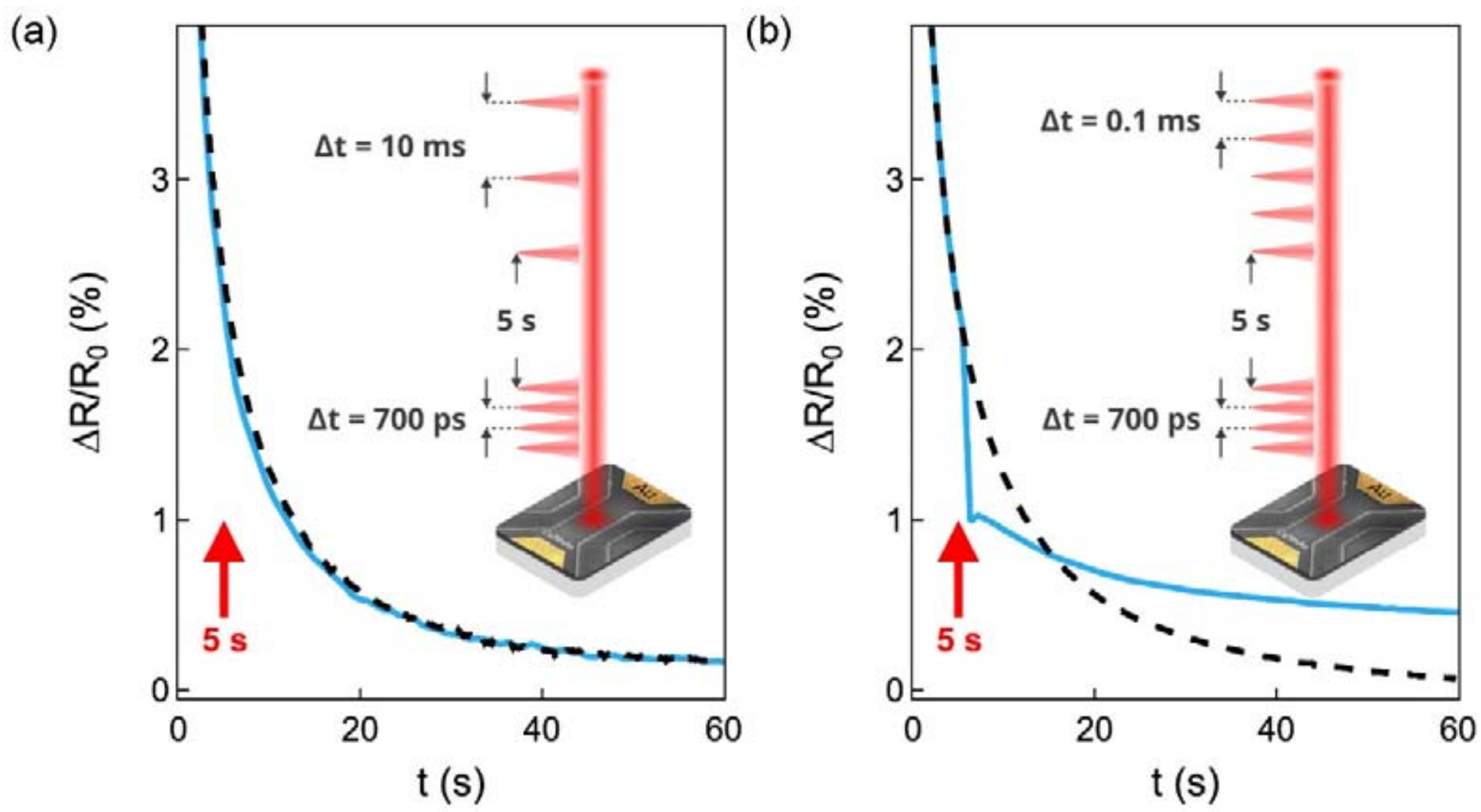

**Supplementary Figure S5. Experimentally-measured dependence of laser-induced erasing of LTM on time spacing between the erasing pulses**. The dashed curve shows the resistivity decay (i.e., a LTM "forgetting" dynamic) measured after a 10 ns long memorizing period, when 16 rehearsals by pulses with a fluence of 0.4 $F_{TH}$ and a time spacing of 700 ps were used in a device fabricated from a 20-nm-thick film. The solid curve in **a,** and **b,** shows the forgetting dynamics when at 5 s laser pulses with the same fluence but with a time spacing of 10 ms and 0.1 ms, respectively, start to be incident on the memory device. The results for a time spacing of 1 ms are shown in Fig. 4d of the main paper.

To further clarify the boundaries between the write, erase, and inert regimes a simple case study of a temperature-induced acceleration of the post-switching CuMnAs relaxation was computed. see Fig. S6. As discussed in detail in Ref. S2, the decay of the high-resistive quench-switched state is described by a Kohlrausch (stretched-exponential) form with $\beta$ = 0.6 and two, strongly temperature-dependent, relaxation times. At room temperature, these times are in the ~10 ms and ~10 s ranges [S2]. On the seconds-long time window relevant to Fig. 4d in the main paper and Supplementary Fig. S5, the fast component is already relaxed and the time evolution of the switching-induced decay of the resistance change can be modeled as

$$\frac{\Delta R}{R_0} \sim exp\left(-\left(\frac{t}{\tau_{QS}(T)}\right)^{0.6}\right). \tag{S5.1}$$

The temperature ($T$) dependence of the characteristic time constant $\tau_{QS}$ follows a simple exponential (Arrhenius) behavior

$$\tau_{QS}(T) = \tau_0 \, exp\left(\frac{E_{QS}}{k_B T}\right), \tag{S5.2}$$

where $\tau_0$ is the attempt time, $E_{QS}$ is the activation energy and $k_B$ is the Boltzmann constant. Using $\tau_0$ = 1 ps and $E_{QS}$ = 0.8 eV [S2], we can reproduce the experimentally measured (a dashed line in Fig. S5) device resistance relaxation to the equilibrium value, as shown by the dashed line in Fig. S6. When the sample temperature is increased by the laser pulse train, which starts to be incident on the device 5 s after the memorizing period, $\tau_{QS}$ is reduced (see inset in Fig. S6b) and the resistance relaxation is accelerated, as shown in Fig. S6 for three values of the laser-induced temperature increase $\Delta T$. Moreover, when $\Delta T$ exceeds ≈ 10 K, also the temperature dependence of CuMnAs resistence (Supplementary Fig. S4a in Ref. S3) has to be taken into account, as shown in Fig. S6c. Finally, we can compare the modelled resistance relaxations with those measured for different pulse spacing $\Delta t$. For $\Delta t$ = 10 ms (Fig. S5a) the resistance relaxation is not visibly accelerated by the laser pulse train, implying $\Delta T < 1$ K (Fig. S6a). For $\Delta t$ = 1 ms (Fig. 4d in the main paper) and 0.1 ms (Fig. S6b) the laser-induced temperature increase seems to be around 5 K (Fig. S6b) and 15 K (Fig. S6c), respectively. This shows that the characteristic time constant of the heat dissipation, which is responsible for this laser-induced erasing of LTM due to the sample temperature build-up (analogous to Fig. S4), is in the milisecond time range.

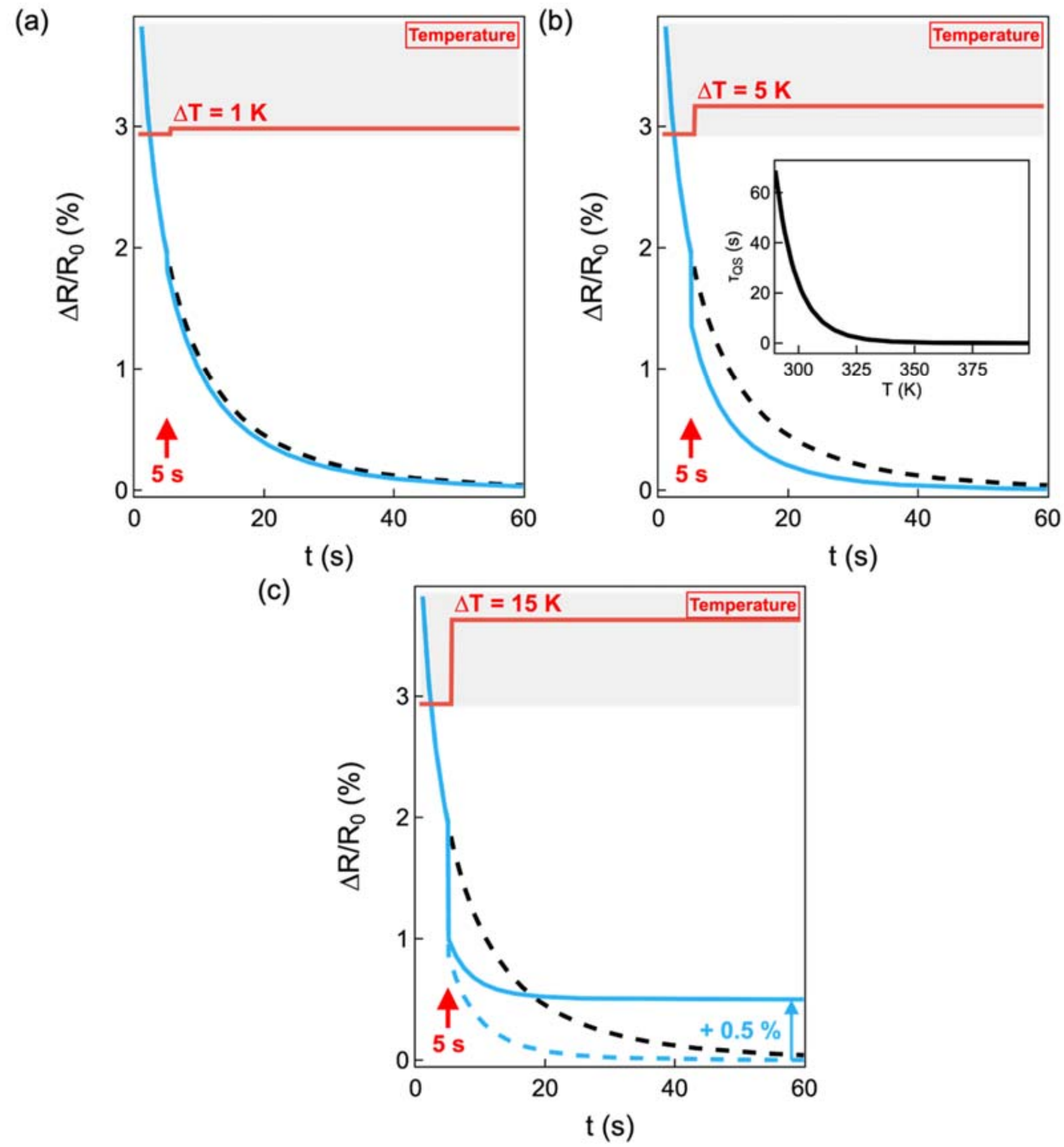

**Supplementary Figure S6. Modelling of temperature-induced erasing of LTM.** Black dashed lines are the room temperature resistance relaxation following the information writing in LTM, as modelled by Eq. (S5.1). When the CuMnAs temperature is increased at $t$ = 5 s by a train of laser pulses (see the upper parts of the figures), the time constant $\tau_{QS}$ is reduced, as shown in the inset in **b**. **a**-**c**, Resistance dynamics (blue lines) obtained for depicted temperature profiles with $\Delta T$ = 1 K, 5 K, and 15 K, respectively. In **c**, due to a considerable temperature increase of 15 K, the resistance does not relax to the equilibrium value (blue dotted line) but to a resistance increased by ≈ 0.5% (see the temperature dependence of $R_0$ in Supplementary Fig. S4a in Ref. S3).